# Theoretical frameworks for neuroeconomics of intertemporal choice.

Taiki Takahashi[1]

[1]Direct all correspondence to Taiki Takahashi, Unit of Cognitive and Behavioral Sciences, Department of Life Sciences, School of Arts and Sciences, The University of Tokyo, Komaba, Meguro-ku, Tokyo, 153-8902, Japan (taikitakahashi@gmail.com).

**Acknowledgements:** The author thanks Ms. Joanna R. Schug at Hokkaido University for critical reading of the manuscript. The research reported in this paper was supported by a grant from the Grant- in-Aid for Scientific Research ("21st Century Center of Excellence" Grant) from the Ministry of Education, Culture, Sports, Science and Technology of Japan.

**Theoretical frameworks for neuroeconomics of intertemporal choice.**

Abstract:
Intertemporal choice has drawn attention in behavioral economics, econophysics, and neuroeconomics. Recent studies in mainstream economics have mainly focused on inconsistency in intertemporal choice (dynamic inconsistency); while impulsivity/impatience in intertemporal choice has been extensively studied in behavioral economics of addiction. However, recent advances in neuroeconomic and econophysical studies on intertemporal choice have made it possible to study both impulsivity and inconsistency in intertemporal choice within a unified framework. In this paper I propose the new frameworks for investigations into neuroeconomics of intertemporal choice. The importance of studying neurochemical and neuroendocrinological modulations of intertemporal choice and time-perception (e.g. serotonin, dopamine, cortisol, testosterone, and epinephrine) is emphasized.

Keywords: *Econophysics*; *Impulsivity*; *Intertemporal choice*; *Inconsistency*; *Neuroeconomics*; *Neuroendocrinology*

**Extended Abstract**:

Intertemporal choice has drawn attention in behavioral economics, econophysics, and neuroeconomics and under strong influences by neuropharmacological modulations. This confirms the importance of studying intertemporal choice by utilizing neuroeconomic methodology. Consider the following intertemporal choice problems:

(A) Choose between (A.1) One cup of coffee now.
(A.2) Two cups of coffee tomorrow.
(B) Choose between (B.1) One cup of coffee in one year.
(B.2) Two cups of coffee in [one year plus one day].

Most people may make the impulsive choice in problem (A) (i.e., choosing (A.1), preference for the small immediate reward); while the same subjects make the patient choice in problem (B) (i.e., choosing (B.2), preference for the large delayed reward). The combination of these typical intertemporal choices is inconsistent, because the time-interval [between (A1) and (A2)] and the time-interval [between (B1) and (B2)] are the same (=1 day). Most people's intertemporal choice *plan* in the example B will be reversed in the intertemporal choice *action* in the example A, as the time of executing the planned action approaches to the present.

Although decision under uncertainty has been well formulated with the prospect theory, behavioral economic theories on intertemporal choice still have problems in explaining observed anomalies (e.g., dynamic inconsistency). Contrary to normative economic theory on intertemporal choice (i.e., exponential discounting), it has now been well established that human and animal intertemporal choice behaviors are not rational (i.e., inconsistent). For this reason, recent economic theory on dynamic optimization and behavioral decision theory on intertemporal choice have adopted a *hyperbolic* discount model, rather than an *exponential* discount model. The hyperbolic discount model can explain the widely-observed tendency of human and animal intertemporal choice---i.e., decreasing impatience. In other words, subjects overestimate their patience in the distant future, resulting in preference reversal as time passes. The preference reversal in hyperbolic discounting may explain various problematic behaviors by humans---loss of self-control, failure in planned abstinence from addictive drugs, or a deadline rush due to procrastination.

Neurobiological and psychological factors determining individual differences in intertemporal choice have been explored in recent behavioral and neuro- economics.

Recent studies in mainstream economics have mainly focused on inconsistency in intertemporal choice (dynamic inconsistency) especially in terms of economic policy; while impulsivity/impatience in intertemporal choice has been extensively studied in the behavioral economics of addiction,. Behavioral and neuroeconomic studies have found that (i) addicts (e.g., cigarette smokers, alcoholics, pathological gamblers, heroin addicts, cocaine addicts, and amphetamine abusers) have large time-discount rates in comparison to non-drug-using controls, (ii) neuroactive hormones such as cortisol (a stress hormone) and testosterone (a male hormone) are associated with temporal discounting, (iii) neurotransmitters such as serotonin, dopamine, and adrenaline (or adrenalin/epinephrine, of which activity in the brain can be non-invasively be assessed by measuring salivary alpha-amylase levels), and related neural circuits, are associated with temporal discount rates. Therefore, while it can be said that the ways in which impulsivity in intertemporal choice (indicated by a discount rate) is modulated by neurobiological factors has intensively studied, another important factor in intertemporal choice, (i.e., inconsistency in intertemporal choice) has been less extensively examined in a quantitatively rigorous manner in neuroeconomics.

Fortunately however, recent advances in neuroeconomic and econophysical studies on intertemporal choice have made it possible to dissociate impulsivity and inconsistency in intertemporal choice within a unified framework. More tellingly, in the q-exponential discount model (based on non-extensive thermostatistical physics), impulsivity and consistency are distinctly parametrized. The q-exponential discount model is capable of parametrizing subject's inconsistency and impulsivity, separately, in a continuous manner. On the other hand, the generalized quasi-hyperbolic discount model, recently proposed in neuroeconomics---note that the conventional quasi-hyperbolic model (i.e., a $\beta-\delta$ model) has been proposed in behavioral economics---can parametrize an internal conflict within an agent between impulsive and patient selves at each time-point in intertemporal choice. However, the relationships between these two models and conventional exponential and hyperbolic discount models have not been studied in neuroeconomics. I stress the importance of examining neuromodulation of the parameters in these two novel discount models in future neuropsychoeconomic studies.

In this paper, I will first review the current status of behavioral and neuro-economics of temporal discounting, with an emphasis on the neurochemistry of intertemporal choice. The effects of neurotranstmitters such as serotnonin and dopamine, and neuroactive hormones such as stress hormones, male hormones, and adrenaline (epinephrine) on human intertemporal choice are introduced. I next propose new theoretical frameworks for investigations into the neuroeconomics of intertemporal choice, and show how these novel frameworks may lead to a better understanding of neuropsychoeconomic decision processes in intertemporal choice. The role of nonlinear time-perception following the Weber-Fechner law in intertemporal choice is also stated. The psychophysics of time-perception may explain many types of anomalies in intertemporal choice----hyperbolicity, subadditivity, and delay/date effects. Finally, some conclusions and implications from the present analysis are presented in relation to future study directions. In discussing future study directions, I have also mentioned the importance of examining the relationship between intertemporal choice and decision under uncertainty, and the usefulness of the novel frameworks for analyzing consistency of governmental economic policy-making.

In summary, this article addresses the importance of studying neurochemical and neuroendocrinological modulations of time-perception and intertemporal choice (e.g. serotonin, dopamine, cortisol, testosterone, adrenaline), as well as the extension of the analysis into group and governmental decision processes by utilizing the present proposed theoretical frameworks. The study directions suggested in this article may aid in the understanding of neurobiological processing underlying intertemporal choice, and resolve the libertarian paternalism controversy occurring between behavioral and neoclassical economists.

## 1. **Intertemporal choice by humans and animals**

Subjects often prefer small but immediate rewards to large but delayed ones (temporal discounting). Behavioral and neuro- economic studies have demonstrated that, in human and animal discounting behavior, “impulsivity” in intertemporal choice (“impatience”) is dependent on the length of delay until receipt (Ainslie, 2005; Bickel and Marsch, 2001; Bickel et al., 2006; Frederick et al., 2002; Mazur, 1987; Prelec, 2004; Reynolds, 2006; Takahashi, 2005, for a formal model of discounting in neoclassical economics, see Appendix I). Specifically, when choosing between smaller sooner rewards and larger later ones, people tend to be patient in the distant future but impulsive in the near future. In order to illustrate how this "decreasing impatience" is associated with inconsistency in intertemporal choice (referred to as "preference reversal" and "dynamic inconsistency" in economics), let us consider the following two intertemporal choice problems:

(A) Choose between (A.1) One cup of coffee now.
(A.2) Two cups of coffee tomorrow.
(B) Choose between (B.1) One cup of coffee in one year.
(B.2) Two cups of coffee in [one year plus one day].

Most people may make the impulsive choice in problem (A) (i.e., choosing (A.1), preference for the small immediate reward); while the same subjects make the patient choice in problem (B) (i.e., choosing (B.2), preference for the larger but more delayed reward). The combination of these typical intertemporal choices is inconsistent, because the time-intervals between [from (A1) to (A2)] and [from (B1) to (B2)] are the same (=1 day). As the time of executing the planned action approaches to the present, most people's intertemporal choice *plan* in example B will be reversed in the intertemporal choice *action* in example A. Therefore, it can be said that most people are patient in their *plan* for the distant future, but impulsive in their intertemporal choice *action* occurring in the near future ("decreasing impatience"), resulting in a “preference reversal” as time passes. This time-inconsistent intertemporal choice cannot be described with the exponential discount function (Mazur, 1987; Prelec, 2004; Ainslie, 2005). Furthermore, this "dynamic inconsistency" is also problematic for governmental policy making, as well as for individual decision-making. Accordingly, several economists have been awarded the Nobel Prize in economics for their contributions to understandings of dynamic inconsistency (i.e., Finn E. Kydland and Edward C. Prescott in 2004; Edmund S. Phelps in 2006).

Because animals also show this type of time-inconsistent intertemporal choice behavior, several animal learning theorists proposed a (simple) *hyperbolic* discount model: $V(D)=V(0)/(1+kD)$, where V(D) is the subjective value of the reward available at delay D (V(0) is therefore the subjective value at delay D=0, i.e., the value of an immediate reward) (Ainslie, 2005; Mazur, 1987). It is to be noted that larger $k$ values indicate more rapid (steeper) discounting. In hyperbolic discounting, the basic assumption of the discounted utility model is that the subjective value of a series of delayed rewards is equivalent to the sum of the instantaneous discounted utilities, but the discount rate is time-dependent. Specifically, for hyperbolic discounting, the discount rate:= $-(dV(D)/dD)/V(D)=k/(1+kD)$ (see Appendix I, for the general definition of a discount rate), which is a decreasing function of delay D (note that in usual settings, the discount rate is positive: $k>0$, referred to as "positive time preference"). A number of behavioral economic, neuropsychopharmacological studies have reported that human and animal intertemporal choice behavior is better described by the hyperbolic model, when compared to the exponential model (Ainslie, 2005; Bickel and Marsch, 2001; Crean et al., 2002; Estle et al., 2007; Frederick et al., 2002; Kable and Glimcher, 2007; Mazur, 1987; Ohmura et al., 2005; Ohmura et al., 2006; Reynolds et al., 2003; Reynolds and Schiffbauer, 2004; Reynolds, 2006; Sozou, 1998; Takahashi, 2005; Takahashi et al., 2007e). Let us examine the neuropsychoeconomic implications of inconsistent hyperbolic discounting behavior.

First, it is to be noted that the preference for more immediate rewards *per se* is not irrational or inconsistent, because there are opportunity costs and risk associated with non-gaining (e.g., due to possible hazards during waiting for the delayed rewards) in delaying the rewards. Therefore, impulsivity in intertemporal choice (a large discount rate) is rationalizable for several types of people. In Becker and Murphy's theory of rational addiction (1988), addicts are supposed to have large discount rates, leading to ignoring future delayed health loss and preferring immediate euphoria obtained from drug intake, in a completely consistent (i.e., rational) manner. This behavior is clinically problematic, but economically rational when their choices are time-consistent (i.e., if they have large discount rates with an exponential discount function). However, it is known that addicts also discount delayed outcomes hyperbolically (Bickel and Marsh, 2001; Bickel et al., 2006; Reynolds, 2006, for a review), suggesting the intertemporal choices of addicts are time-inconsistent, resulting in a loss of self-control (see Bickel and Marsch, 2001, for a review). It is to be stressed here that problem B is a *plan* regarding intertemporal choice in the distant future; i.e., at one year later from now (because the agents cannot take future actions now); while problem A is an intertemporal choice *action*. The time-inconsistency between intertemporal choice *plan*s and *action*s is problematic in the sense that even if an agent had made patient and forward-looking plans regarding the distant future (as in problem B), as the time of executing the plan approaches the present s/he will cancel the patient plan and act more impulsively at the moment of intertemporal choice *action*s, against his/her own previously-intended plan. In other words, there is a discrepancy between the decision-maker's intentions (will) and actions (behavior), indicating that in intertemporal choice problems, most people cannot act as they intended/planned in advance (Ainslie, 2005; Frederick et al., 2002; Bickel and Marsch, 2001; Takahashi, 2005).

A number of economic studies (e.g., Laibson, 1997) state that this dynamic inconsistency may explain various problematic behaviors. The problematic behaviors associated with inconsistency in intertemporal choice which have been studied are: loss of self-control, failure in formerly-planned abstinence from addictive substances and relapse, a dead-line rush due to procrastination, failure in saving enough before retirement (although people at younger ages tend to plan/want to save for their retirement in later life, most of them at middle ages do not reduce consumptions enough), and risky sexual behavior (Chesson et al., 2006).

In addition to the "dynamic inconsistency", previous behavioral economic studies have observed several anomalies in intertemporal choice; *e.g.*, sign effects (gain is more rapidly discounted than loss), and magnitude effects (smaller gain is more rapidly discounted than larger gain) (see Bickel and Marsch, 2001; Frederick et al., 2002, for a review). However, while neurochemical studies have examined neuropharmacological modulations of these effects in intertemporal choice (Bickel and Marsch, 2001; Takahashi et al., 2006 b; Takahashi et al., 2007 c; Takahashi et al., 2007 g), no neuroimaging studies have investigated the neural processing underlying these anomalies (but also see Berns et al., 2006, for a neuroimaging study of intertemporal choice for pain). More neuroimaging studies are needed for elucidating neuropsychoeconomic correlates of these anomalies. Furthermore, a "domain effect" in intertemporal choice (i.e., directly consumable rewards such as foods and addictive substances are more rapidly discounted than money) has also been observed (Bickel et al., 1999; Estle et al., 2007). A recent neuroimaging study (McClure et al., 2007) examined the neural correlates of temporal discounting for a primary reward (i.e., water) and reported that the reward-processing neural circuits for delay discounting of money and water were distinct. It should further be noted that “actually experience now" and "plans of an imagined future" may induce distinct neuropsychological processes in temporal discounting. Specifically, the difference between choices A and B may be fundamental (in spite of the same time-intervals between the choices 1 and 2) in the sense that in A you have the possibility to consume/experience the reward now, that is, to give in to the urge for coffee now. This speculation is in line with the domain effect. Future neuroeconomic studies should examine the neural correlates of temporal discounting for addictive substances which subjects are dependent on.

Because information regarding the neurochemical substrates underlying intertemporal choice is important for (i) establishing neuropharmacological treatment of addiction and (ii) elucidating intertemporal choice behavior at the molecular level, I will next introduce neurochemical substrates for computational models of reinforcement learning and intertemporal choice in relation to brain regions (Section 2). Then, I will explain novel mathematical frameworks for temporal discounting behavior which have been recently introduced in neuroeconomics (Section 3). Section 4 presents more detailed characteristics of the novel discount models, with emphasis on the time-dependency of discount rates and behavioral interpretations. Furthermore, I will denote a hypothetical account of time-inconsistency in intertemporal choice based on the psychophysics of time-perception (Section 5). Finally, I will discuss some conclusions and implications with proposals for specific research directions based on the unified view (Section 6).

## 2. **Neurochemistry of intertemporal choice**

Several neurobiologists and economists have recently started to examine neural correlates of intertemporal choice by utilizing neuroimaging techniques (McClure et al., 2004; Hariri et al., 2006; McClure et al., 2007; Monterosso et al., 2007; Wittmann et al., 2007a; Kable and Glimcher, 2007). Until recently, neurobiological studies on temporal discounting have mainly been conducted by computational neuroscientists in order to model reinforcement learning processes, especially in temporal difference (TD) learning theory (Dayan and Abbott, 2001), and parameters in the theory have been associated with neurochemical substrates (Yu and Dayan, 2005; Schultz, 2004; Schweighofer et al., 2007).

"Reinforcement learning" has been investigated extensively in computer science because degree of computational complexity and memory loading of the reinforcement learning algorithms are lower than that of conventional dynamic programming algorithms (Bellman, 1957; Sutton and Barto, 1998), and therefore it is plausible that the reinforcement learning algorithms may be implemented in the brain (Dayan and Abbott, 2001). The central objective of reinforcement learning is to estimate the value function (i.e., a summation of subjective values of delayed rewards):

$$V(s(t))=\mathrm{E}[\sum_{k=0}^{\infty}\delta^{k}v(t+k)], \qquad \text{(Equation 2-1)}$$

where $v(t)$, $v(t+1)$, $v(t+2)$, … are the rewards acquired by following a certain policy $P(a|s)$ (defined below) starting from state $s(t)$, and $\delta$ is a discount factor independent of time $t$ (such that $0 \le \delta \le 1$). Note that E[x] indicates the expected value of x, following the standard notation in statistics. Therefore, the reinforcement theory adopts the discrete-time exponential discounting framework. The value functions for the states before and after the state-transition should satisfy the recursive relation (i.e., a consistenty equation, Sutton and Barto, 1998):

$$V(s(t-1))=\mathrm{E}[v(t)+\delta V(s(t))]. \qquad \text{(Equation 2-2)}$$

Hence, the deviation (i.e., the prediction error) from the consistency equation

$$\Delta(t)=v(t)+\delta V(s(t))-V(s(t-1)) \qquad \text{(Equation 2-3)}$$

should be zero on average. This deviation is defined as the temporal difference (TD) error in reward prediction (Sutton and Barto, 1998; Dayan and Abbott, 2001) and is utilized to update the value function:

$$\Delta V(s(t-1))=\alpha\,\Delta(\tau). \qquad \text{(Equation 2-4)}$$

where $\alpha$ is the learning rate (Sutton and Barto, 1998). The "policy" function is often defined via the action value function Q(s(t),a) which represents the value of future rewards the subject would obtain by taking the action $a$ at the state $s(t)$. The policy function defined in this manner is expressed as:

$$P(a_i|s(t))=\frac{\exp(\beta_L Q(s(t),a_i))}{\sum_{j=1}^{M}\exp(\beta_L Q(s(t),a_j))}, \qquad \text{(Equation 2-5)}$$

where the parameter $\beta_L$ is called the "inverse temperature" (because this parameter plays the same role as the inverse of absolute temperature in the Boltzmann distribution function utilized in thermostatistical physics, Dayan and Abbott, 2001). We here assume the subject utilizes the "soft-max" strategy, following the standard theory of reinforcement learning in theoretical neuroscience (Dayan and Abbott, 2001). It is to be noted that the efficiency of the reinforcement learning theory crucially depends on parameter values of $\alpha$, $\beta_L$ and $\delta$. It is also important to note that $\delta$ corresponds to a time-constant discount factor (see Appendix I, for the relation between a discount rate and a discount factor), indicating that agents following the reinforcement learning theory discount delayed rewards in a time-consistent manner and never experience "preference reversal" over time. Neurochemical studies on the reinforcement learning theory have proposed that $\alpha$, $\beta_L$ and $\delta$ are associated with distinct neurotransmitters, i.e., acetylcholine, norepinephrine (noradrenaline), and serotonin, respectively. *In vivo* neurophysiological studies in primates have established that dopaminergic neural activities of the striatum in the brain encode the TD error signals in the reinforcement learning theory (Dayan and Abbott, 2001; Schultz, 2004). Therefore, it is important to examine neuromodulation of temporal discounting in order to further develop the reinforcement learning models in neuroscience. It is also again to be noted that, contrary to the assumption in the reinforcement learning theory, behavioral studies have reported that humans and animals discount rewards hyperbolically (Mazur, 1987; Ainslie, 2005; see Frederick et al., 2002, for a review, but see Schweighofer et al., 2006).

Several neuroeconomic studies have reported that neurochemical substrates are associated with human and animal subjects' hyperbolic discount rates (again note that most behavioral studies observed that subjects discount hyperbolically, rather than exponentially; in contrast to computational models of reinforcement learning). These investigations into the neurochemical substrates underlying hyperbolic discounting are important for neuroeconomic understandings of impulsivity and inconsistency in intertemporal choice, aas well as for reducing these problematic tendencies. It is to be noted that impulsive intertemporal choice is clinically problematic (especially in drug addicts); while inconsistency in intertemporal choice is both clinically and economically problematic. The advantages of studying the neurochemical correlates of intertemporal choice is that (i) these studies may reveal rigorous biophysical mechanisms underlying hyperbolic discounting and (ii) neurochemical understanding of hyperbolic discounting may help to establish medical tretments (e.g., pharmacological treatment) for addiction and problematic economic behaviors. However, it must be noted that because we still do not have non-invasive methods for selectively activating and inactivating neural activities in humans, from an ethical perspective, psychopharmacological treatments are the only available neurobiological methods in humans. To date, several types of neuromolulators such as serotonin, dopamine, addictive drugs such as nicotine and heroin, as well as neuroactive hormones such as cortisol, testosterone, and adrenaline have been reported to be associated with intertemporal choice. Because most studies on the neurochemical modulation of intertemporal choice assume the hyperbolic discount model and examine how the discount rate of the hyperbolic model is associated with the concentration of neurochemical substances and the activities of neuromodulators, we now focus on the neuromodulations of impulsivity in temporal discounting. The problem of ignoring neuropharmacological correlates of the inconsistency in temporal discounting in most literature in neuropsychopharmacological studies of intertemporal choice is discussed later (see Section 3).

**2.1 Serotonin and intertemporal choice**

Serotonergic systems are known to relate to mood disorders such as depression, and anti-depressants (such as *prozac*, a type of SSRIs, selective seronotnin reuptake inhibitors) typically activate serotonergic neural activities by reducing the reuptake of serotonin in synaptic extracellular spaces in neural circuits (Asberg et al., 1986). Mobini et al (2000a) was the first study to report that a reduction in serotonergic activities increases hyperbolic discount rates in rodents. The same authors reported that the alterations in serotonergic systems did not affect risk attitude in decision under uncertainty ("probability discounting") in rodents (Mobini et al., 2000b). Therefore, it can be concluded that serotonergic activities in the rodent brain modulate intertemporal choice, but not risk-taking behavior. In humans however, a reduction in serotonergic activities did not significantly increase a time-discount rate (Crean et al., 2002; Schweighofer et al., 2006), indicating that the role of serotonergic activity in decision-making is still inconclusive. It is important to further investigate the roles of serotonergic systems in intertemporal choice in order to understand impulsive behaviors observed in psychiatric patients (e.g. suicide attempts by depressive patients, Renaud et al., 2007). Moreover, a recent study by computational neuroscientists has demonstrated, based on the TD model in reinforcement learning theory, that the modulation of serotonergic activities induced by the regulation of tryptophan (a precursor of serotonin) levels affected the evaluation of delayed reward in a time-dependent manner (Tanaka et al., 2007).

**2.2 Dopamine and intertemporal choice**

Dopaminergic neural systems (e.g., the ventral tegmental area, the nucleus accumbens) are reward-processing brain regions. Recent developments in reinforcement learning theory of dopaminergic systems have elucidated the importance of reward prediction (Dayan and Abbott, 2001; Schultz, 2004). Specifically, neurons in the striatum encode the reward predictin error in TD model (the deviation from the consistency equation) in the reinforcement learning theory; while the neural activity of nucleus accumbens encodes the prediction/anticipation of the value of a reward (Dayan and Abbott, 2001; Schultz, 2004). The dopaminergic systems also have pivotal roles in temporal discounting. Notably, behavioral economist Loewenstein and cognitive neuroscientist Cohen's group has reported that impulsive intertemporal choice between immediate and delayed money is associated with the activation of dopaminergic systems such as the ventral tegmental area (McClure et al., 2004); and Kable and Glimcher (2007) demonstrated that a subjective value of a delayed reward is associated with the activation of dopaminergic neural circuits. Also, it is known that a reduction in a serotonergic activity in the brain attenuates the reducing effect of dopaminergic drugs (*d*-amphetamine) on hyperbolic discount rates (Winstanley et al., 2005), indicating that there are interactions between serotonergic and dopaminergic systems in modulating intertemporal choice. Furthermore, it has been proposed that electrical coupling via non-synaptic gap junctions between dopamine neurons may cause inconsistency in intertemporal choice via altered time-perception (Takahashi, 2005; Takahashi, 2006a). Therefore, dopmaninergic modulations of intertemporal choice and time-perception should be more extensively examined in future neuroeconomic studies (a mathematical model relating time-perception to intertemporal choice will be introduced later). Regarding neuropsychiatric illnesses, a recent study reported that schizophrenic patients (who are known to have altered dopamine activities) have larger discount rates in comparison to healthy controls (Heerey et al., 2007).

## 2.4 Addictive drugs and intertemporal choice

Chronic intake of addictive drugs such as nicotine, heroin, (metha)amphetamine and cocaine has been associated with impulsive intertemporal choice (nicotine and discounting: Bickel et al., 1999; Ohmura et al., 2005, Reynolds et al., 2003; heroin and discounting: Kirby et al., 1999; Bretteville-Jensen, 1999; amphetamine and discounting: Bretteville-Jensen, 1999; Hoffman et al., 2006). These findings are consistent with Becker and Murphy's theory of rational addiction (Becker and Murphy, 1988), in that addicts put small weight on their health values in later life. The next question is whether these substance abusers and addicts were originally impulsive in intertemporal choice (before the onset of drug addiction) or have become impulsive due to the neuropsychopharmacological effects of habitual drug intake. Recent studies have investigated this question by employing human and animal subjects. In human studies, it is ethically problematic to administrate addictive drugs to non-drug-dependent healthy subjects. Therefore, recent studies have examined the stability of addicts' discount rates over time after abstinence. If large discount rates are due to habitual drug intake, it is expected that discount rates decreased after long-term abstinence. However, it has recently been reported that for alcoholics and smokers, abstinence did not dramatically reduce discount rates of former alcoholics and smokers (Yoon et al., 2007; Takahashi et al., 2007c).

In animal studies, Dallery and Locey (2005) have reported that chronic nicotine administration causes long-lasting increases in time-discount rates in rats, and Simon et al (2007) reported the same effect of cocaine in rats. Together, it can be supposed that the chronic intake of addictive drugs may increase discount rates in a long-lasting manner. However, no study to date has examined whether addictive drugs increase time-inconsistency in intertemporal choice.

### 2.5 Neuroactive hormones and intertemporal choice

It is well known that human and animal behaviors are under strong influences of neuroactive hormones such as cortisol (a human stress hormone), testosterone (a male hormone), and estradiol (a female hormone). The roles of the hormones in human social behavior have extensively been examined; while those in economic decision-making have relatively less extensively been studied (Caldú and Dreher, 2007). These hormones are the neuromodulators of the dopaminergic systems, and control the saliency of rewards via the effects on alertness, motivation, and arousal (Kelly et al., 2005). Furthoremore, electrophysiological studies demonstrated that dopaminergic, reward-processing neurons have receptors of a stress hormone and their activity was modulated by a stress hormone via glucocorticoid receptors (Saal et al., 2003). Namely, addictive drugs (e.g., ethanol, cocaine, and nicotine) and stress hormones trigger a common neuronal adaptation in dopaminergic circuits at the molecular level (Kauer, 2003). It is therefore expected that intertemporal choice is also under the influences of the neuroactive hormones.

The first investigation into the roles of neuroactive steroid hormones has been performed on cortisol (a stress hormone) in humans. Cortisol is secreted via the activation of hypothalamic-pituitary-adrenal (HPA) axis in response to stressors (McEwen, 2003) and has euphoric and anxiolytic effects appearing via the activations of glucocorticoid receptors in the brain, although cortisol also has memory impairment effects in the hippocampus (Takahashi et al., 2004 b). Drug addicts have low arousal and therefore reduced cortisol response. This reduction in cortisol response and arousal in addicts may be associated with their drug intake, because cortisol has both euphoric and arousing effects (Plihal et al., 1996). Consistent with these findings, Takahashi (2004 a) demonstrated that subjects with low cortisol levels were more impulsive in intertemporal choice, indicating that deficiency in cortisol-induced euphoric actions in the brain under stressful events may lead to a demand for the immediate pleasure available from drug intake. This may result in the observation that drug addicts have low cortisol levels and attenuated cortisol response to psychosocial stressors (al'Absi, 2006). This is also in line with the van Honk et al.(2003)'s report that impulsive and disadvantageous behavior in the Iowa gamgling task (referred to as "future myopia" and "insensitivity to future consequences", Bechara et al., 1994) was associated with low cortisol levels.

Testosterone, a male hormone, has been associated with impulsivity and aggression in males (Archer, 2006). However, examinations into the relationships between testosterone and temporal discounting are quite few. Takahashi et al. (2006) reported that when hyperbolic discount *rate* was utilized for assessing the impulsivity in intertemporal choice, an inverted U relationship between discount rate and testosterone level was observed in male humans. Also, Takahashi (2007) reported that, when a discount *factor* is utilized as a measure of patience in intertemporal choice, testosterone was positively associated with male subjects' discount factors. These results may reflect complex interactions between testosterone and dopaminergic activities, and conversion of testosterone into estradiol (a female hormone) in the brain (Hojo et al., 2004). In any case, testosterone was not related to impulsivity in temporal discounting for losses. This indicates that anti-androgen therapy may not be effective when the subjects' problematic behavior is associated with insensitivities to future bad outcomes.

Because (a) adrenaline (or adrenalin/epinephrine) is associated with alertness and arousal (Ramos and Arnsten, 2007) and (b) patience and self-control in intertemporal choice requires high degrees of arousal, (note that sleep deprivation increases discount rates, Reynolds and Schiffbauer, 2004), the relationship between adrenergic activities and temporal discounting may be important. However, only one study to date has examined this relationship (Takahashi et al., 2007 d). Because adrenaline levels in the brain cannot directly be measured in a non-invasive manner, Takahashi et al (2007 d) has utilized salivary alpha-amylase levels as an index of adrenergic activities. They demonstrated that subjects with lower salivary alpha-amylase levels were more impulsive in intertemporal choice (a larger hyperbolic discount rate), in line with the previous finding that subjects with low arousal (e.g., sensation seekers) are more impulsive and susceptible to drug addiction (Zuckerman, 1990).

Collectively, it can be said that impulsivity in temporal discounting is strongly influenced by addictive drugs and neurochemical substrates including neuroactive hormones. This is consistent with the proposed neuromodulation of parameters in reinforcement learning theory. However, as we have seen, little is known regarding the neuromodulation of dynamic inconsistency in intertemporal choice. As noted, the reinforcement learning theory assumes exponential discounting, rather than hyperbolic discounting (dynamically-inconsistent discounting), and it is still unknown how hyperbolic discounting should be implemented in the reinforcement learning theory. Furthermore, there has been no good measure of inconsistency in intertemporal choice in behavioral studies (but see Prelec, 2004, for an axiomatic definition of time-inconsistency, i.e., "decreasing impatience"). Strictly speaking, it cannot be concluded that the mentioned neuromodulations of the degrees to which subjects discount delayed rewards are solely attributable to changes in discount rates. Specifically, it is possible that the neuromodulators had also changed the functional forms of subject's temporal discounting function. There has been no appropriate framework for assessing the degree of hyperbolicity of the time-discounting function (i.e., time-inconsistency). Recent studies in econophysics and neuroeconomics have developed better discount models; i.e., the q-exponential discount function and the generalized quasi-hyperbolic discount function. I will introduce these models below and discuss the usefulness of these new discount models in neuropsychoeconomics.

## 3. Novel models of temporal discounting behavior

Recently, behavioral neuroeconomic and econophysical studies have proposed two discount models, in order to better describe the neural and behavioral correlates of impulsivity and inconsistency in intertemporal choice. It is to be noted that "econophysics" is an interdisciplinary field in which theoretical tools of statistical physics are applied to economic and financial phenomena (see Stanley et al., 1996; Stauffer, 2004, for a review). Econophysical studies have traditionally paid attention to problems in macroeconomic and financial phenomena (e.g., income inequality, inefficiency in stock markets), rather than those in individual decision-making. Recently however, individual choice such as decision under risk has also been attracting econophysicists' attention (Bordley, 2005). In econophysical studies on intertemporal choice, it may be useful to employ mathematical frameworks of Tsallis non-extensive thermostatistics (Tsallis, 1988) which have originally been applied to study non-classical physical characteristics of stellar structures (Plastino and Plastino, 1993). Actually, Tsallis statistics-based frameworks have also been utilized for elucidating human decision-making, perception of risk, and biological systems (Tsallis, 1995).

### 3.1 q-exponential discount model

As described above, neuroeconomic studies of addiction have been focusing on impulsivity in intertemporal choice, rather than dynamic inconsistency, although the inconsistency in intertemporal choice is also important especially in order to better understand addicts' problematic behavior (e.g., relapse). In order to describe human and animal subject's intertemporal choice behaviors in a manner which we can dissociate impulsivity and inconsistency, recent econophysical studies (Cajueiro, 2006; Takahashi et al, 2007 e) have proposed and examined the following q-exponential discount function for subjective value V(D) of delayed reward:

$$V(D)=A/\ exp_q(k_qD)=A/[1+(1-q)k_qD]^{1/(1-q)} \qquad \text{(Equation 3.1)}$$

where $exp_q\ (.\ )$ is a "q-exponential" function (formally defined below), D is a delay until receipt of a reward, $A$ is the value of a reward at $D$=0, and $k_q$ is a parameter of impulsivity at delay $D$=0 (q-exponential discount rate). The q-exponential function is defined as:

$$exp_q(x):=[1+(1-q)x]^{1/(1-q)}.$$

We can easily see that this q-exponential function approaches a usual exponential function in the limit of $q \rightarrow 1$ (by utilizing l'Hospital's theorem or the definition of an exponential function). The q-exponential function has been extensively utilized in studies in econophysics in the application of Tsallis' non-extensive thermostatistics which may explain income distributions following power functions (Michael and Johnson, 2003), although the q-exponential discount model has not frequently been utilized in neuroeconomics.

It is to be noted that when $q$=0, equation 1 is the same as a hyperbolic discount function (i.e., $V(D)=A/(1+k_qD)$), while $q \rightarrow 1$, is the same as an exponential discount function (i.e., $V(D)=Aexp(-k_qD)$). In exponential discounting (q→1 in equation 1), preference reversal never occurs, because the discount rate:= $-(dV/D)/V=k_q$ is time-independent when q→1. While the dynamics of discount rates will more extensively be addressed below, the important point here is that the q-exponential discount model can distinctly parametrize impulsivity and dynamic consistency in intertemproal choice. Conventional models of temporal discounting (i.e., exponential and hyperbolic discount models) cannot achieve this economically important goal. Furthermore, because econophysics is also a new important approach to understanding paradoxes in economic and financial phenomena, econophysic frameworks may aid in solving problems in neuroeconomics. However, very rarely have neuroeconomics and econophysics been combined.

Takahashi et al. (2007 e) have shown that the q-exponential discount function is capable of continuously quantifying human subjects' inconsistency in intertemporal choice. Namely, human agents with smaller q values are more inconsitent in intertemporal choice. If $q$ is less than 0, the intertemproal choice behavior is more inconsistent than hyperbolic discounting (in other words, the discount rate of the q-exponential function with $q<0$ more rapidly decreases than that of the simple hyperbolic discount function, i.e., “hyper-hyperbolic”, see 4.1, for mathematical details). Additionally, $1\text{-}q$ can be utilized as an inconsistency parameter. Hence, future studies should examine neuromodulation of $k_q$ (impulsivity) and q (dynamic consistency) in the q-exponential discount model. An important application of the q-exponential discount model may be to examine whether addicts are more inconsistent than non-drug-dependent controls. This examination may be a direct test of the rational addiction theory (Becker and Murphy, 1988), because Becker and Murphy's rational addiction theory states that addicts consistently maximizes the sum of discounted utilities from drug intake (immediate reward) and health loss (delayed loss). The consistency here refers to exponential discounting (Becker and Murphy, 1988). Recently, health economists have examined whether addicts are more time-inconsistent than healthy controls (Blondel et al., 2007), but this study has limitations in the sense that the analytical framework was not quantitatively rigorous. The present framework may provide a more rigorous methodology of assessing addicts' consistency in intertemporal choice.

### 3.2 generalized quasi-hyperbolic discount model

Behavioral economists have proposed that the inconsistency in intertemporal choice may be attributable to an internal conflict between "multiple selves" within a decision maker (Laibson 1997; O'Donoghue and Rabin, 1999). The proposal is persuasive in the sense that we often feel regret after an impulsive choice. The existence of the feeling of regret even after acting as one intended (i.e, not being forced to make a certain choice) may be a strong evidence of human behavior's inconsistency and internal conflict between desires inconsistent with each other. This hypothesis states that (a) there are (at least) two exponential discounting selves (i.e., two exponential discount rates) in a single human individual and (b) when delayed rewards are at the distant future (>1 year), the self with a smaller discount rate wins; while delayed rewards approach to the near future (within a year), the self with a larger discount rate wins, resulting in preference reversal over time (Laibson 1997; O'Donoghue and Rabin, 1999). This intertemporal choice behavior has been referred to as quasi-hyperbolic discounting (also as a $\beta$-$\delta$ model, see below) (Laibson 1997; O'Donoghue and Rabin, 1999). Originally, the quasi-hyperbolic model has been proposed in a discrete-time version. In the discrete time, the quasi-hyperbolic discounting $F(\tau)$ for discrete time $\tau$ (the unit has been assumed to be one year) is defined as (Laibson, 1997):

$F(\tau)=\beta\,\delta^{\tau}$ (for $\tau$=1,2,3,…) and F(0) =1 ($0<\beta<\delta<1$).

The quasi-hyperbolic discount model captures human bias in intertemporal choice; *i.e.*, a discount factor between the present and one-time period later (β) is smaller than that between two future time-periods (δ). In other words, people are patient in planning their intertemporal choice in the distant future, but impulsive in intertemporal choice action occurring at delay D=0. Therefore, this simple quasi-hyperbolic function has a kink at the time point of τ=1 (year). This model therefore predicts a distinction in impulsivities between an intertemporal choice executed at delays >1 year and <1 year. A previous neuroeconomic study (McClure et al., 2004) demonstrated that the activation in dopaminergic midbrain regions (the ventral striatum) was associated with β; while the activation in the prefrontal cortex was associated with δ, indicating that there is a "neural conflict" between impulsive and patient selves in the brain.

Furthermore, a recent neuroeconomic study on temporal discounting for primary rewards in thirsty subjects utilizing functional magnetic resonance neuroimaging (McClure et al., 2007), has proposed a generalized quasi-hyperbolic discount model in which "dual selves" are linearly weighted at each delay. In the continuous time, the study's proposed model is equivalent to the linearly-weighted two-exponential functions (generalized quasi-hyperbolic discounting):

$$V(D)=A[w\,exp(-k_1D)+(1-w)exp(-k_2D)] \qquad \text{(Equation 3.2)}$$

where $0<w<1$ is a weighting parameter and $k_1$ and $k_2$ are two exponential discount rates ($k_1<k_2$). Note that the larger exponential discount rate of the two ($k_2$) corresponds to an impulsive self; while the smaller discount rate $k_1$ corresponds to a patient self. Also, it is easy to see that if w=1 or 0, this discount model is the same as the exponential discounting. The mentioned neuroeconomic study, employing thirsty human subjects, observed that, within a single subject, the "impulsive self" (a larger discount rate $k_2$) was associated with neural activity in the reward-processing midbrain regions; while the "patient self" ($k_1$) was associated with prefrontal brain regions in temporal discounting for primary rewards (McClure et al., 2007). However, it must be noted that in this neuroimaging study, the generalized quasi-hyperbolic model was originally proposed as a discrete-time discount function. Although the psychological interpretation of inconsistent intertemporal choice behavior in quasi-hyperbolic discounting is intuitively appealing, no study to date has compared the goodness-of-fit between the (generalized) quasi-hyperbolic discount model and the q-exponential discount model in an empirical manner. This examination should be performed in future neuropsychoeconomic studies.

## 4. Discount rates in q-exponential and generalized quasi-hyperbolic discounting

Because explicit expressions of a discount rate and its time-dependency are important, next I will summarize the properties of discount rates in the q-exponential and the generalized quasi-hyperbolic discount models. It is important to again note that in any continuous time-discounting function, a discount rate is defined as $-(dV(D)/dD)/V(D)$, independently of functional forms of discount models, and larger discount rate indicates more impulsive intertemporal choices (see Appendix I).

### 4.1 Discount rate in the q-exponential discount model

In the q-exponential discount model, the discount rate as defined is:

$$\text{(q-exponential discount rate)}=k_q/(1+k_q(1-q)D). \qquad \text{(Equation 3.3)}$$

We can see that when $q$=1, the discount rate is independent of delay $D$, corresponding to exponential discount model (consistent intertemporal choice); while for $q$<1, the discount rate is a decreasing function of delay $D$, resulting in preference reversal. This can be seen by a direct calculation of the time-derivative of the q-exponential discount rate:

$$(d/dD)\text{(q-exponential discount rate)}= -k_q^2(1-q)/(k_q(1-q)D+1)^2 \qquad \text{(Equation 3.4)}$$

which is negative for $q$<1, indicating "decreasing impatience" for q smaller than 1. Also, impulsivity at delay D=0 is equal to $k_q$ irrespective of $q$. Therefore, $k_q$ and $q$ can parametrize impulsivity and consistency, respectively, in a distinct manner. Notably, when q is negative, the speed of a decrease in the q-exponential discount rate is faster than the hyperbolic discount rate (i.e., “hyper-hyperbolic”). Because this distinction between impulsivity and dynamic inconsistency is important for economic theory, future neuroeconomic studies should examine how neuromodulators such as serotonin, dopamine, addictive drugs, and neuroactive hormones affect $k_q$ and $q$, distinctly.

### 4.2 Discount rate in the generalized quasi-hyperbolic discount model

For the generalized quasi-hyperbolic discount model (Equation 2), the discount rate is:

$$[k_2(1-w)exp(-k_2D)+k_1w\ exp(-k_1D)]/[(1-w)exp(-k_2D)+w\ exp(-k_1D)]. \qquad \text{(Equation 3.5)}$$

It is to be noted that at delay $D$=0, the generalized quasi-hyperbolic discount rate = $wk_1+(1-w)k_2$. This indicates that impulsivity in an intertemporal choice *action* (at delay D=0) corresponds to linearly-weighted discount rates at delay D. The time-derivative of the discount rate, i.e., ($d/dD$) (a generalized quasi-hyperbolic discount rate) is:

$$-w(1-w)(k_2-k_1)^2 exp(k_2D+k_1D)/[wexp(k_2D)+(1-w)exp(k_1D)]^2, \qquad \text{(Equation 3.6)}$$

which is negative because 1-$w$>0. This also indicates that the discount rate is a decreasing function of delay, again indicating "decreasing impatience". Neuropsychologically, the weighted difference between discount rates $k_2>k_1$:

$$(1-w)k_2-wk_1 \qquad \text{(Equation 3.7)}$$

may indicate the strength of "internal conflict" between impulsive and patient selves in intertemporal choice, and this can be regarded as an internal conflict parameter. The reason is that this parameter indicates the degree of disagreement between weighted impulsive and patient selves in intertemporal choice. No neuroeconomic study to date has examined the relationship between the conflict parameter and the q-parameter in the q-exponential discount model, i.e., an inconsistency parameter. One promising direction of future neuroeconomics studies may be to examine the relation between the conflict parameter $(1-w)k_2-wk_1$ and inconsistency parameter (1-$q$) in the two models in combination with neuroimaging.

## 5. Time-perception and intertemporal choice

Recent behavioral, neuro- economic, and neuropharmacological studies collectively stress the importance of time-perception in intertemporal choice. It is to be noted that in the history of decision theory on uncertainty, both utility functions and probability have been replaced with more (neuro)psychological quantities, i.e., a value function with a gain-loss asymmetry and probability weighting function with a distortion in probability perception in prospect theory (Kahneman and Tversky, 1979; Prelec, 1998). More specifically, in the classical expected utility theory, the subjective value of an uncertain reward (each outcome $x_i$ has a probability $p_i$) has the form: $U(x_i, p_i)=\Sigma_i p_i u(x_i)$; while in the present prospect theory, the subjective value ("prospect") is $\Sigma_i w(p_i)v(x_i)$ where w(p) is a probability weighting function and $v(x_i)$ is a value function with a gain-loss asymmetry. Therefore, it may be supposed that the intertemporal choice theories will also experience similar changes. Loewenstein and Prelec (1992) took an initial step towards this direction, by attempting to explain the sign effect in intertemporal choice; i.e., gains are more rapidly discounted than losses of the same magnitudes (Frederick et al., 2002). According to this theory, the perceived magnitude of \$10 loss is larger than that of \$10 gain (because of the gain-loss asymmetry in the value function). Consequently, \$10 loss is less rapidly discounted than \$10 gain due to the magnitude effect (i.e., large outcomes are less rapidly discounted than smaller outcomes). In the Loewenstein-Prelec theory, a subjective value of a delayed reward is in the form of $v(x)F(D)$ where v(x) is a value function (rather than a utility function) and $F(D)$ is a discount function at delay D, which is still a purely economic quantity. Similar to prospect theory, it can be hypothesized that the correct discount function has the form: $F(\tau(D))$, where $\tau(D)$ is a subjectively perceived length of delay D.

Notably, Takahashi (2005) has proposed that exponential discounting with logarithmic time-perception: $\tau(D)=\alpha \log(1+\beta D)$ may explain dynamic inconsistency in intertemporal choice. More tellingly, if a subject discounts a delayed reward exponentially, but with the logarithmic time-perception (i.e., Weber-Fechner law in psychophysics), his/her temporal discount function has a hyperbolic form: $F(\tau)=exp(-k\ \tau)=1/(1+\beta D)^{k\alpha}$. Intuitively speaking, subjects try to discount a delayed reward exponentially (i.e., rationally and consistently), but actual intertemporal choice behavior may be hyperbolic and dynamically inconsistent, due to a distortion in time-perception (It is also to be noted that the exponential discount model with logarithmic time-peception is mathematically equivalent to the q-exponential discount model based on Tsallis' statistics). There is accumulating evidence to support this hypothesis. First, one neuroimaging study has observed that when subjects make intertemporal choices, brain regions for time-perception such as the caudate nucleus (a type of dopamine systems) are activated in association with the delay length (Wittmann et al., 2007 a). Secondly, subjects with large discount rates have overestimated time-perception. For instance, substance abusers (Wittmann et al., 2007 b) and sleep-deprived subjects (Reynolds and Schiffbauer, 2004), known to have large discount rates, have prolonged time-perception. Thirdly, behavioral economists have reported that if the time of receiving a delayed reward is presented in the form of a calendar date (instead of time durations of delay length until receipt) the functional form of their temporal discounting becomes exponential, rather than hyperbolic (Read et al., 2005). This "delay/date effect" can be explained by considering that the presentation of a calendar date may reduce the non-linearity of the perception of delay length. Moreover, subadditive time-discounting---i.e., a discount factor is a decreasing function of a time-interval between two alternatives---may also be explained by non-linear psychophysical transformation of objective time into psychological time (Read, 2001; Takahashi, 2006 a). Together, it appears to be promising to examine the relationships between neuropharmacological modulation of time-perception and that of intertemporal choice. Future neuropsychoeconomic studies along this line may help to understand the neural processing underlying the dynamic inconsistency in interetemporal choice.

## 6. **Conclusions and implications**

The present theoretical frameworks and proposals for future neuroeconomic studies on intertemporal choice have several implications: (i) one of the most important topics in economics, i.e., dynamic inconsistency, is still an unresolved difficult problem which should be attacked by neuroeconomists, (ii) investigations into neurochemical modulations of both impulsivity and inconsistency in intertemporal choice may be promising clues for better understanding the neural bases of intertemporal choice and this direction may help establish clinical treatment for problematic behaviors such as addiction, (iii) novel discount models such as the q-exponential discount model and the generalized quasi-hyperbolic model may be helpful in examining dynamic inconsistency in intertemporal choice, and (iv) the relationship between time-perception and intertemporal choice should more extensively studied by utilizing the present theoretical frameworks via neurochemical manipulation and neuroimaging. These future directions can probably be explored through standard neurobiological methods such as non-invasive neuroimaging with functional magnetic resonance imaging, neuropsychopharmacological manipulations, and neurogenetic analysis based on a huge amount of human genome data.

With respect to the better understanding of neuro-cognitive processing involved in intertemporal choice, it may be important to examine the relationship between intertemporal choice and decision under risk (Lee, 2005). Several studies have proposed that a framework for intertemporal choice can also be utilized for decision under uncertainty (Rachlin, 1991; Takahashi et al., 2007 f). Notably, Rachlin et al (1991) have proposed that risk-aversion is attributable to temporal discounting (based on a molar maximization theory which assumes ergodicity in stochastic processes); while other studies try to attribute temporal discounting to risk-aversion and explain the hyperbolicity in temporal discounting (Sozou, 1998). Further studies are apparently necessary in order to reconcile these hypotheses.

Moreover, Rachlin and colleagues have demonstrated that generosity/altruism for other people decreases as social distances for the others increase in a hyperbolic manner (Jones and Rachlin, 2006). Therefore, it is of potential interest to examine whether common neurocognitive processing is underlying both temporal and social discounting (Takahashi, 2007 a). Specifically, if people both logarithmically perceive delay and social distance, as well as try to discount the values of a reward with the perceived delay and social distance, the reported hyperbolicity of both temporal and social discount functions may be attributable to a common psychophysical law (i.e. Weber-Fechner law).

Furthermore, the utilization of the present novel frameworks (i.e., the q-exponential discount model and the generalized quasi-hyperbolic model) may also prove useful in analyzing governmental economic policy-making. The well-known behavioral economist Richard Thaler and his colleagues have proposed that there is a need for "libertarian paternalistic" policies for correcting biases in individual's behavior (Sunstein and Thaler, 2003), based on recent accumulating behavioral evidence that human behavior deviates from rational/normative economic theory. These behavioral anomalies include hyperbolic discounting, dynamic inconsistency, myopic loss aversion, and irrational addiction. However, these inconsistent decision-making manners may also be problematic in governmental policy-making (Garfinkel and Lee, 2000). This point is under strong controversy between behavioral economists such as Richard Thaler and neoclassical economists such as Gary Becker (Becker, 2006). Therefore, future neuropsychoeconomic studies should examine whether dynamic inconsistency in intertemporal choice is mitigated or not when the decision was made by other people (e.g., governmental officials), in order to resolve the controversy and establish ways to make more efficient socioeconomic policies (including drug regulation policies, see Badger et al., 2007, for an economic study on the related issue). The first attempt has already been performed: it was observed that intertemporal decision-making for others is more inconsistent and impulsive than decisions made for decision-maker herself (Takahashi, 2007 b). Combining present neuroeconomic theories with group decision theories may also help to resolve these issues.

Finally, I will now propose suggestions for future studies based on the unification of the present frameworks. Specific examples may better illustrate how the present frameworks can be unified for elucidating neuropsychoeconomic processing underlying intertemporal choice.

First, it may be important to examine (through fMRI etc.) the neural correlates of impulsivity ($k_q$ in the q-exponential discount function), time-consistency ($q$ in the q-exponential function), and the degree of conflict between selves ($(1-w)k_2-wk_1$ in the generalized quasi-hyperbolic discount function) in intertemporal choice by addicts (e.g., alcoholics, smokers, and heroin abusers), by utilizing the present frameworks. Our present theoretical examination predicts that (a) time-consistency and the degree of conflict may negatively be correlated, (b) addicts have larger impulsivity at delay=0 ($k_q$) than non-drug-dependent controls, (c) $k_q$ may be positively associated with the activation of the midbrain dopaminergic systems (e.g., the ventral striatum), and (d) the conflict parameter $(1-w)k_2-wk_1$ may be associated with the activation of the anterior cingulated cortex (a conflict monitoring neural circuit) and the difference in the activations between the prefrontal and midbrain regions. It would also be important to examine the ways in which the q-exponential discount model can be incorporated into the reinforcement learning theory, since the reinforcement learning theory has extensively been utilized in primate neurophysiology and human fMRI studies (Dayan and Abbott, 2001; Schultz, 2004; Tanaka et al., 2007). It is also important to incorporate the q-exponential disconting into reinforcement learning theory, because actual intertemporal choice behavior is not exponential (assumed in the reinforcement learning theoretical models, Dayan and Abbott, 2001), but q-exponential (or hyper-hyperbolic). This investigation can be combined with neuroimaging studies. Moreover, as noted, the exponential discounting with logarithmic time-perception is equivalent to the q-exponential discounting. Therefore, how the psychophysical and neuropsychopharmacological effects on time-perception affect parameters of the q-exponential function should be examined.

Additionally, in terms of economic policy-making it is of interest to compare the intertemporal choices for made for one's self versus choices made for someone else via neuroimaing. The unified consideration from our present frameworks predicts that (a) during intertemporal choice for self, the midbrain dopaminergic regions may be more strongly activated than during that for someone else (because the receipt of monetary gain may be more rewarding than the giving), (b) simple hyperbolic and exponential discount rates may be smaller for intertemporal choice for others rather than the self, because the discount rate of intertemporal decisions for someone else more rapidly decreases than that for self (i.e., intertemporal choice *plan* for someone else is less impulsive than that for self), although $k_q$ (the q-exponential discount rate), namely impulsivity at delay D=0 (impulsivity in intertemporal *action*) is larger for someone else than that for self, (c) both prefrontal and midbrain activations during intertemporal choice for someone else may be smaller than those for self. Future neuroimaging studies may examine whether the predictions for the examples above are correct or not.

## Appendix I:
## Intertemporal choice in neoclassical microeconomics

Most people prefer an immediate reward to a delayed reward in intertemporal choice (delay discounting) (Ainslie 2005; Frederick et al., 2002; Ohmura et al., 2006). Intertemporal choice behavior is related to addiction, saving behavior, financial planning such as investment and retirement. Therefore, studies in behavioral and neuro-economics, psychopharmacology, behavioral finance, and econophysics have focused on how intertemporal choice is characterized in theoretical and empirical manners. In game theory, discounting is also important for understanding how iterated versions of games such as the prisoner's dilemma and self-control in intertemporal choice may be a neuropsychological constraints on the evolution of cooperation (Stevens and Hauser, 2004), consistent with the folk theorem in game theory. In this section, a conventional framework for the economic modeling of intertemporal choice is introduced according to behavioral economist Laibson's article (Laibson, 2003).

In neoclassical mainstream microeconomics, intertemporal choice has been formulated through the discounted utility model proposed by Paul Samuelson (see Frederick et al., 2002, for a review). In the discounted utility model, it is assumed that a consumer's welfare can be represented as a discounted sum of current and future instantaneous utility at each time point. Suppose that, at each time, the decision-maker consumes $c(t)$. The instantaneous subjective value of the consumption is a utility function $u(c(t))$.

In continuous-time intertemporal choice, the sum of the instantaneous discounted utilities is:

$$\int_0^D F(D)u(c(t+D))dD$$

where F(D) is a discount function at delay D. A discount rate at delay D is defined as:

$$r(D) := -(dF(D)/dD)/F(D).$$

Note that larger discount rates correspond to more impulsive (less patient) intertemporal choice. On the other hand, a discount factor f(D) at delay D is defined as:

$$f(D) := \lim_{\Delta \to 0} \left(\frac{1}{1+r(D)\Delta}\right)^{\left(\frac{1}{\Delta}\right)} = \exp(-r(D))$$

Note that smaller discount factors indicate more impulsive intertemporal choice. In conventional neoclassical economics, it has been assumed that the discount rate is independent of D, confirming dynamic consistency of intertemporal choice. In this case, let us put $r(D)=\rho$, which is independent of delay D. The exponential discount function is then $F(D)=exp(-\rho D)$. This exponential discount function can also be expressed as $F(D)=\delta^D$ where $\delta:=$exp(-ρ). If we express the discount rate and the discount factor in terms of δ, we obtain:

$$r(D)= \rho= -ln\delta$$

and

$$f(D)=exp(-\rho)=\delta.$$

Therefore, in exponential discounting in the continuous-time formulation, (discount rate) = $-ln$ (discount factor) and both are independent of delay D. In the Nobel Prize winning economist Becker and his colleague Murphy's theory of rational addiction, it is assumed that addicts maximize the sum of future discounted utilities with an exponential discount function, resulting in completely rational addiction even in the case of the strongest addiction to heroin, alcohol, and nicotine (Becker and Murphy, 1988). It is important to note that a number of neuropsychopharmacological studies have demonstrated that addicts (e.g., heroin addicts, cocaine addicts, alcoholics, and smokers of cigarettes) have larger discount rates, in comparison to healthy non-drug-dependent subjects (Bickel and Marsch, 2001). This may be in line with the economic theory of addiction (Becker and Murphy, 1988), in terms of the degree to which a subject discounts a delayed reward (a discount rate). However, a rational intertemporal choice should have consistency over time (dynamic consistency) and empirical studies have repeatedly observed that human intertemporal choice does not have rationality in this sense.